\documentclass[sigconf,screen=true]{acmart}
\usepackage{listings}
\usepackage{xcolor}
\usepackage{hyperref}
\usepackage{cleveref}
\usepackage{tcolorbox}
\usepackage{subcaption}

 \renewcommand\footnotetextcopyrightpermission[1]{}  

\copyrightyear{2020}
\acmYear{2020}
\setcopyright{acmlicensed}
\acmConference[ESEM '20]{ESEM '20: ACM / IEEE International Symposium on Empirical Software Engineering and Measurement (ESEM)}{October 8--9, 2020}{Bari, Italy}
\acmBooktitle{ESEM '20: ACM / IEEE International Symposium on Empirical Software Engineering and Measurement (ESEM) (ESEM '20), October 8--9, 2020, Bari, Italy}
\acmPrice{15.00}
\acmDOI{10.1145/3382494.3410636}
\acmISBN{978-1-4503-7580-1/20/10}

\definecolor{keywordpurple}{RGB}{128,5,85}
\definecolor{stringblue}{RGB}{42,0,255}
\definecolor{commentblue}{RGB}{95, 143, 191}
\definecolor{numberpink}{RGB}{255, 0, 170}
\lstdefinestyle{javastyle}{
    commentstyle=\color{commentblue},
    keywordstyle=\color{keywordpurple},
    numberstyle=\tiny\color{numberpink},
    stringstyle=\color{stringblue},
    basicstyle=\ttfamily\footnotesize,
    breakatwhitespace=false,
    breaklines=true,                 
    captionpos=b,                    
    keepspaces=true,              
    numbersep=5pt,
    showspaces=false,
    showstringspaces=false,
    showtabs=false,
    tabsize=2
}
\begin{document}

\title{An Empirical Validation of Cognitive Complexity as a Measure of Source Code Understandability}

\author{Marvin Mu\~{n}oz Bar\'{o}n}
\affiliation{%
  \institution{University of Stuttgart}
  \city{Stuttgart}
  \country{Germany}
}
\email{st141535@stud.uni-stuttgart.de}

\author{Marvin Wyrich}
\affiliation{%
  \institution{University of Stuttgart}
  \city{Stuttgart}
  \country{Germany}
}
\email{marvin.wyrich@iste.uni-stuttgart.de}
\orcid{0000-0001-8506-3294}

\author{Stefan Wagner}
\affiliation{%
  \institution{University of Stuttgart}
  \city{Stuttgart}
  \country{Germany}
}
\email{stefan.wagner@iste.uni-stuttgart.de}
\orcid{0000-0001-8506-8429}

\begin{abstract}
  \textbf{Background}: Developers spend a lot of their time on understanding source code. Static code analysis tools can draw attention to code that is difficult for developers to understand. However, most of the findings are based on non-validated metrics, which can lead to confusion and code, that is hard to understand, not being identified.
  
  \noindent\textbf{Aims}: In this work, we validate a metric called \textit{Cognitive Complexity} which was explicitly designed to measure code understandability and which is already widely used due to its integration in well-known static code analysis tools.
  
  \noindent\textbf{Method}: We conducted a systematic literature search to obtain data sets from studies which measured code understandability. This way we obtained about 24,000 understandability evaluations of 427 code snippets. We calculated the correlations of these measurements with the corresponding metric values and statistically summarized the correlation coefficients through a meta-analysis.
  
  \noindent\textbf{Results}: Cognitive Complexity positively correlates with comprehension time and subjective ratings of understandability. The metric showed mixed results for the correlation with the correctness of comprehension tasks and with physiological measures.
  
  \noindent\textbf{Conclusions}: It is the first validated and solely code-based metric which is able to reflect at least some aspects of code understandability. Moreover, due to its methodology, this work shows that code understanding is currently measured in many different ways, which we also do not know how they are related. This makes it difficult to compare the results of individual studies as well as to develop a metric that measures code understanding in all its facets.
  
\end{abstract}


\maketitle

\newpage
\section{Introduction}
\label{sec:introduction}

Understanding source code is an integral part of the software development process.
On average, professional developers spend more than 50\% of their time on activities related to program comprehension~\cite{Minelli.2015,Xia.2018}.
Consequently, it is vital to provide them with sufficient information to assess the understandability of their code and identify potential for improvement.
In a comprehensive study investigating 121 different existing code- and documentation-related metrics as well as metrics relating to a developer's background and experience, Scalabrino et al. found that none significantly correlated with code understandability~\cite{Scalabrino.2019}.
Even though none of them were originally intended to measure understandability, they are commonly assumed to do so.

At the same time, a metric called \textit{Cognitive Complexity} was introduced, which was explicitly designed to measure source code understandability~\cite{Campbell.2018}.
The metric is already being reported to a significant number of developers, as it has become a default measure in the SonarSource\footnote{https://www.sonarsource.com/} tool landscape.
A preliminary evaluation showed that developers accept the metric to the extent that they mostly resolve findings of high Cognitive Complexity in their source code~\cite{Campbell.2018}.
However, it has not yet been evaluated whether or not Cognitive Complexity captures code understandability.

Unfortunately, a lack of empirical evaluation seems to be the reality for most software metrics employed in today's static analysis tools.
Recent reports indicate that out of the hundreds of metrics used for analysis, only as few as twelve are validated in scientific studies~\cite{Nilson.2019}.
Using metrics without proper validation can lead to unsound decisions from developers and maintainers.
The refactoring of already well understandable code is not only time consuming but can also introduce new defects. 
At the same time, the absence of proper detection of incomprehensible pieces of code, prevents a shared understanding of the codebase among developers.

Therefore, we present an empirical validation of Cognitive Complexity as a measure of source code understandability.
Using a systematic literature search, we identified studies that measure code comprehension from a human developer's point of view and built an aggregated data set with data from \textbf{10 studies} spanning over \textbf{427 code snippets} and approximately \textbf{24,000 individual human evaluations}.
Using this data, we performed a comprehensive meta-analysis to investigate the correlation of Cognitive Complexity with measures of source code understandability.

\section{Background}
\label{sec:background}
This section describes what source code understandability is, introduces the central metric evaluated in this paper, and discusses related work that has attempted to correlate software metrics with understandability.

\begin{figure*}[t]
    \begin{subfigure}{.53\textwidth}
\begin{lstlisting}[language=Java]
public static List<Integer> primes(int[] input) { //Cycl Cogn        
    List<Integer> primes = new ArrayList<>();

    for (i : input; i++) {                        // +1   +1
        if(isPrime(i)) {                          // +1   +2
            primes.add(i);
        }
    }
    return primes; }
\end{lstlisting}
    \end{subfigure}
    \begin{subfigure}{.46\textwidth}
\begin{lstlisting}[language=Java]
public static void slots() {                //Cycl Cogn   
    int temp = rand.nextInt(1000); 
    switch(temp)                            //      +1
        case 777:                           // +1
            System.out.println("Jackpot");
        case 123:                           // +1
            result();
        default:                            // +1
            System.out.println("Game Over"); }
\end{lstlisting}
    \end{subfigure}
    \caption{Measuring of Cyclomatic and Cognitive Complexity. (In total: Cyclomatic Complexity=2|3, Cognitive Complexity=3|1)}
    \Description{Two code samples showing that Cognitive and Cyclomatic Complexity assign different scores for the same code structures.}
    \label{fig:compare-cog-cycl}
\end{figure*}

\subsection{Source Code Understandability}
There have been numerous efforts to describe what code understandability is and what other software quality attributes could be considered as its influencing factors.
In essence, understanding source code is to read it and deduce its meaning.
Boehm et al. describe understandability as the extent to which ``code possesses the characteristic of understandability to the extent that its purpose is clear to the inspector''~\cite{Boehm.1976}.
In this work, in particular, we consider bottom-up comprehension, in which the programmer analyzes the source code line-by-line and from several lines, deduces \lq chunks\rq{} of higher abstraction and finally aggregates these chunks into high-level plans~\cite{OBrien.2004}.
This approach stands in contrast to top-down comprehension, which is usually applied when one is already familiar with the code or the type of code~\cite{Mayrhauser.1995}.

Some terms are closely related to understandability, such as readability and complexity.
Boehm et al. state that legibility is necessary for understandability~\cite{Boehm.1976}.
Intuitively, it seems likely that if code is harder to read, it is harder to understand.
But on the other hand, while a piece of code might be considered as readable, developers might still have difficulties understanding it~\cite{Scalabrino.2019}.
Therefore we explicitly consider understandability and readability distinct factors that nevertheless are closely related.
Unfortunately, there is not a field-wide standard and in some other works, the distinction is not as clear.
For example, readability is sometimes described as ``the amount of mental effort required to understand the code''~\cite{Sedano.2016} or ``the judgment about how easy a block of code is to understand''~\cite{Buse.2010}.

In an experimental setting, the understandability of source code has been measured in a plethora of different ways.
First, several different tasks can be performed to find out whether a participant has understood a program.
Such tasks include answering comprehension-related questions~\cite{Dolado.2003, Salvaneschi.2014}, filling out blank program parts~\cite{Borstler.2016}, or more advanced tasks, such as extending or modifying existing code or finding bugs~\cite{Hofmeister.2017}.

Usually, different types of measures are then taken when conducting the experiment to assess the degree with which a program was understood.
Most record whether the comprehension task was completed successfully~\cite{Kasto.2013, Woodfield.1981, Wiedenbeck.1999}.
Others also record the time taken to answer questions or perform a task on the code such as locating or fixing a bug~\cite{Hofmeister.2017, Feigenspan.2011}.
This is not the norm, however, as some studies prescribe the time the subjects have to complete the comprehension task~\cite{WIEDENBECK.1999b, Aljunid.2012, RAMALINGAM.1997}.
A more recent trend in the field of program comprehension research is the usage of physiological measures by employing fMRI scanners~\cite{Peitek.2018, Floyd.2017}, biometrics sensors~\cite{Fritz.2014, Yeh.2017, Fucci.2019} or eye-tracking devices~\cite{Turner.2014, Fritz.2014}.
Lastly, some studies measure \textit{perceived understandability}~\cite{Scalabrino.2019} which we explicitly distinguish from the other measures.
In these studies, participants are asked to rate code according to its understandability.
While both constructs of understandability are of interest to us, it is possible that different factors influence how developers perceive the understandability of a code snippet compared to how they would perform in comprehension tasks regarding said snippet.

\subsection{Software Metrics for Understandability}
Since it would not be feasible in practice to conduct an experiment to assess whether new code is understandable each time, it is of great interest to find a way to automatically measure the understandability of any code snippet.
Previous efforts have been made to validate whether existing software metrics can capture code understandability.

Scalabrino et al.~\cite{Scalabrino.2019} conducted a study where they calculated correlations between 121 metrics and proxy variables for understandability gathered in an experiment with professional developers.
In their experiment, they investigated code-metrics like LOC and Cyclomatic Complexity, documentation-related metrics such as comment readability and metrics relating to a developer's experience and background.
They concluded that none of the investigated metrics could accurately represent code understandability.
Even after repeating the study with an increased sample size, the results did not change.
However, they noted that, although the combination of metrics in the models still did not fully capture the understandability of the source code, it improved their effectiveness.
Trockman et al.~\cite{Trockman.2018} reanalyzed the data set from Scalabrino et al. and put a renewed focus on combined metrics.
They employed different statistical methods and found that code features had a small but significant correlation with understandability.
In conclusion, they suggest that a useful metric for understandability could be created but more data would be needed to confirm that notion.

Peitek et al.~\cite{Peitek.2018} investigated the correlation of software metrics with the brain deactivation strength, used as an indicator of concentration levels, captured while performing comprehension tasks in an fMRI scanner.
They found that for all deactivated areas, higher values for the metric DepDegree and Halstead's measures indicated more concentration.
Lines of code showed the weakest correlations with concentration during comprehension tasks.
For Cyclomatic Complexity, they found that higher values were associated with lower concentration among participants.
They did warn, however, that due to the small sample size and due to the snippets not being designed to answer whether there is a correlation, the results might not necessarily be statistically significant despite high values of correlation for some of the metrics.

Kasto et al.~\cite{Kasto.2013} attempted to find a connection between a multitude of different code metrics and comprehension difficulty by analyzing the student answers of a final exam in an introductory Java programming course.
The comprehension tasks included code tracing tasks, where students figured out the output of the source code by hand and "explain in plain English" questions, where they explained the functionality of the code in words.
They found that some of the investigated metrics, i. e. Cyclomatic Complexity, nested block depth, and two dynamic metrics, correlated significantly with the student performance in code tracing exam questions.
No significant correlation could be found with the explanation questions.
One limitation of their results was that the exam questions contained a low number of program commands and did not exceed one or two methods as they were part of a first-year programming course.
They also noted that depending on the way the comprehension questions are posed, they might test other areas besides programming knowledge.
They mention that, for example, the validity of answers to questions including mathematical concepts or operators might be influenced by the student's mathematical knowledge.

Feigenspan et al.~\cite{Feigenspan.2011} measured the correlation of software measures with program comprehension from the data of maintenance tasks taken in an experiment with graduate students.
The investigated variables for understandability were the correctness and response time of the performed tasks.
They could not find a correlation between software measures and code comprehension.

In summary, most existing studies investigate the correlation of software metrics with proxy variables of understandability in laboratory experiments.
There are major differences in their approaches, especially concerning the understandability measures and the comprehension tasks they employ.
Most results indicate that existing metrics fail to capture understandability.
This, combined with the lack of empirical validation of industry-employed source code metrics in general demonstrates a need for the comprehensive evaluation of more metrics in terms of their ability to measure understandability.

\subsection{Cognitive Complexity}

In 2017, SonarSource introduced Cognitive Complexity~\cite{Campbell.2018} as a new metric for measuring the understandability of any given piece of code.
On the surface, Cognitive Complexity appears to be quite similar to the Cyclomatic Complexity metric by McCabe~\cite{McCabe.1976}.
In fact, Cognitive Complexity was specifically designed to address some of the common critiques and shortcomings of Cyclomatic Complexity such as the nesting problem~\cite{SulemanSarwar.2013} and fill what they call the ``understandability gap''~\cite{Campbell.2018}.
\Cref{fig:compare-cog-cycl} shows how Cognitive Complexity assigns different values to structures such as nested \verb|if| statements and \verb|switch case| structures.

In essence, Cognitive Complexity is built on three basic rules~\cite{Campbell.2018}.
First, ignore structures that allow multiple statements to be shorthanded into one.
This means that there is no increment for a method declaration or null-coalescing operators like \verb|??| in C\# or PHP.
Second, there is an increment for breaks in the linear flow.
This includes structures like loops, conditionals, gotos, catch statements, sequences of logical operators and recursive methods.
Third, there is an increment for nested control flow structures.
For example, a nested loop declaration would increase the complexity value by two, an additional structure nested within this loop would increase the value by three and so on.

The initial paper~\cite{Campbell.2018} included a small investigation of the developers' reaction to the introduction of Cognitive Complexity in the static code analysis tool service SonarCloud.
In an analysis of 22 open-source projects, they assessed whether a development team 'accepted' the metric based on whether they fixed code areas of high Cognitive Complexity as reported by the tool.
They found that the metric had a 77\% acceptance rate among developers.
They pose that ``for a metric formulated to bridge Cyclomatic Complexity's understandability gap, the most important measure of success must be developer response''~\cite{Campbell.2018}.
However, they also acknowledge that more studies are needed to assess the validity and utility of Cognitive Complexity.
We agree with their sentiment and do not consider this to be a sufficient validation of Cognitive Complexity since its primary goal of capturing understandability has not yet been evaluated.

\section{Methods}
\label{sec:methods}
Based on the preliminary research, we found a need for an evaluation of the Cognitive Complexity on its merits as a measure of code understandability.
We formulated our research question as follows:
\textit{\textbf{RQ1}: Does Cognitive Complexity correlate with measures of source code understandability from existing studies?}

To validate Cognitive Complexity as a measure of source code understandability, we chose a three-step approach.
First, we conducted a literature search to find data sets from studies that measure the understandability of source code from the perspective of a human developer.
We then filtered them with regards to whether or not we could gain access to their source code and experimental data.
Finally, we used these data sets to investigate whether there was a correlation between the Cognitive Complexity of source code and measures of understandability through a meta-analysis.

\subsection{Systematic Literature Search}
While carrying out our systematic literature search, we mainly followed the guidelines described by Kitchenham et al.~\cite{Kitchenham.2007} for performing systematic literature reviews in software engineering.
The main difference between our approach and the one described by Kitchenham et al. was that our goal was to identify potential data sets published alongside experiments rather than conducting a systematic review of the studies.
While we describe the general search strategy and briefly summarize the results here, more detailed information can be found in the supplemental materials.
There, the papers excluded in each step of the systematic search, technical details on how it was performed, and which tools were used are available.

\newpage
\subsubsection{Search Strategy}

\paragraph{Search Terms}
The search string should return all studies on the subject of code understandability.
To construct it, we included understandability, its synonyms, and some other closely related terms and made sure that the results would relate to software engineering and program source code.
The search string was as follows:
\begin{tcolorbox}[top=3pt,bottom=2pt,left=2pt,right=2pt]
\ttfamily\texttt{(code OR software OR program) AND \\
(understandability 	OR comprehension OR readability \\
OR complexity OR analyzability OR \textquotedbl cognitive load\textquotedbl )
}
\end{tcolorbox}
The first part of the string, \verb|(code OR software OR program)|, was used to make sure that the results would relate to software engineering and program source code.
This qualifier was then combined with understandability, its synonyms and other terms that were closely related to it.

\paragraph{Data Sources}
The aim of gathering a sufficient amount of data sets from papers warranted the use of multiple online data sources.
Google Scholar\footnote{http://scholar.google.com/}, the ACM Digital Library\footnote{https://dl.acm.org/}, IEEE Xplore\footnote{https://ieeexplore.ieee.org/} and Elsevier ScienceDirect\footnote{https://www.sciencedirect.com/} were used to search for literature.

\paragraph{Search Period}
All works published between 2010 and July 2019, the date of the literature search, were included in the search results.
This was done because our initial search terms were very broad and we wanted to limit the number of irrelevant papers that had to go through the manual filtering process.
Since the main objective of this literature search was to find data sets from understandability studies that could be used in the data analysis, a special focus was placed on identifying studies with open data sets.
Open Science has been a trend for some years now, but only recently has it gained momentum.
Due to this development, it made sense to limit the search period to more recent years, as the studies published then were most likely to have openly published their data sets.

While this reduced the number of studies to be checked against the criteria, it also meant that some relevant studies might have been excluded even though they would meet all other criteria.
To mitigate this effect, a backward snowballing step was added to the search strategy, which ignored the search period restriction.

\paragraph{Inclusion and Exclusion Criteria}
From this point on, a more fine-grained approach was used to filter and obtain appropriate studies.
The search results were constrained to peer-reviewed journals, conferences and workshops, as we suspected that this was where the majority of the experiments were reported and where we could find published data to be used in our analysis.
A summary of the criteria used to filter the retrieved literature from the search results is given in \cref{fig:inclusion-exclusion-criteria}.
\begin{table}[h]
	\centering
	\caption{Inclusion and exclusion criteria}
	\Description{Table showing a list of bullet points with the inclusion and exclusion criteria used in the literature search}
	\label{fig:inclusion-exclusion-criteria}
\begin{tcolorbox}[top=3pt,bottom=2pt,left=2pt,right=2pt]
	\textbf{Inclusion criteria}  \\
	\textbullet\ \hspace{.1mm} Title matches the search string \\
	\textbullet\ \hspace{.5mm} Published in a peer-reviewed journal, conference or workshop \\
	\textbullet\ \hspace{.1mm} Published after 2010 \\
	\textbullet\ \hspace{.1mm} Reports on an experiment measuring the understandability of code from a human perspective
	\tcblower
	\textbf{Exclusion criteria} \\
	\textbullet\ \hspace{.1mm} Paper is not related to the field of software engineering \\
	\textbullet\ \hspace{.1mm} Paper is a duplicate entry \\
	\textbullet\ \hspace{.1mm} Paper is not available in English \\
	\textbullet\ \hspace{.1mm} Full text is not available 
\end{tcolorbox}
\end{table}

\paragraph{Snowballing}
Kitchenham et al.~\cite{Kitchenham.2007} suggest looking for additional papers in the reference lists of relevant primary studies, a process commonly referred to as snowballing.
New papers are extracted from the references given in papers that were previously marked as relevant.
Here, a single iteration of backward snowballing was applied to the papers remaining after going through all filtering steps as it is described by Wohlin~\cite{Wohlin.2014}.
Each reference then was put through the same filtering steps of duplicate removal and inclusion/exclusion criteria, except for the time period cutoff.
Special importance was put on this snowballing step since the initial search restricted the results to the years 2010 and later.
This way, the studies conducted before the time cutoff were still included in the final search results as long as they were mentioned in one of the identified studies.

\paragraph{Data Extraction}
At this point, it was assumed that all papers on relevant source code understandability studies were found.
The full text of each paper was then explored to identify whether they had a publicly published data set.
Where this was not the case, but the study mentioned appeared to be of value to the analysis, the authors were contacted to request access to the code snippets and experimental data.
All papers for which we could not obtain a data set were then removed.
For the remaining papers, we then extracted information on the study methodology, the programming language, number of participants, demographic as well as the employed comprehension tasks and measures.
This information can be seen in \cref{table:data-sets-overview}.
Lastly, we followed the references to the published data sets given by the authors and retrieved the source code snippets and experimental data to be used in the analysis.

The experimental data consisted of tables with the identification of the code snippets and the understandability measures that were obtained during the experiments.
Depending on the study, these measures either were the time taken to comprehend a snippet (time), the degree to which a comprehension task on the snippet was completed (correctness), a subjective rating of the understandability of the snippet (rating) and physiological data gathered during the comprehension of a snippet (physiological).
Which measures were taken by each of the studies can be found in \cref{table:data-sets-overview}.

\begin{figure*}[t]
	\centering
    \includegraphics[width=\textwidth]{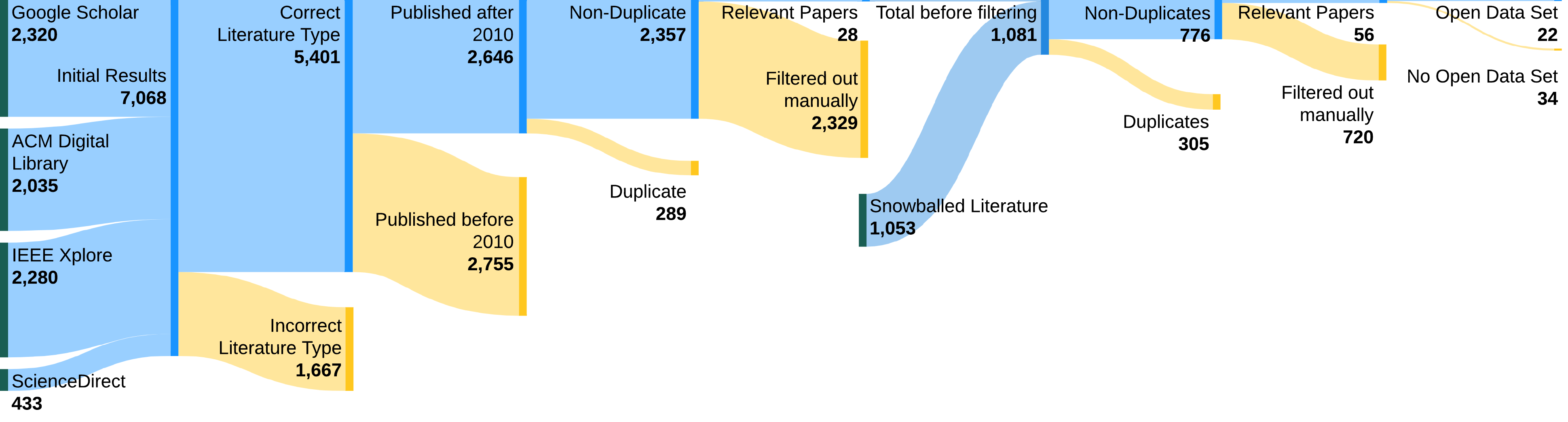}
	\caption{Number of literature search results included and excluded in each step}
	\label{fig:sankey-diagram-sls}
	\Description{Sankey diagram showing how many papers where excluded in each step of the literature search}
\end{figure*}

\paragraph{Data Synthesis}
The data synthesis step differed from the usual systematic literature review process due to the nature of the search results.
Instead of just using the full-text to review the study methodology or subjects, the actual data sets resulting from these studies were the main objective of this literature search.
In addition, the code snippets had to be modified so that they could be compiled and the Cognitive Complexity metric could be measured automatically as described in \cref{data-preparation}.
The methodology of the subsequent analysis can be found in \cref{sec:data-analysis} and the results in \cref{sec:results}.

\subsubsection{Search Results}

\Cref{fig:sankey-diagram-sls} shows the data flow of papers that were analyzed during the filtering stages of the literature search with the number of excluded literature annotated for each step.
Without any filters applied, most literature was found through Google Scholar, with the ACM Digital Library and IEEE Xplore closely following.
Only Elsevier ScienceDirect showed a significantly smaller number of relevant results, only making up about 6\% of the total amount.
In total, out of the initial 7,068 papers, 28 were deemed suitable for this study.
The biggest filter was the restriction on the time period for publishing.
Even with all the automatic filtering steps removing about 4,711 entries, the number of papers that had to be manually checked for suitability was still significant (2,357).

Snowballing proved to be an effective tool to widen the focus of the search, raising the number of relevant papers from 28 to 56.
The majority of the new papers, 20 out of 28, were published before 2010.
Of the 56 relevant papers, only 22 published their experimental data, source code, or a replication package.
We contacted 12 authors of relevant papers that did not publish their data alongside their study.
Of the six researchers that replied, three offered to share their source code and data with us.
The number of data sets we could use ended up being lower than initially thought, as eight papers reported on the same four studies, leading to only four unique data sets.
Additionally, we had to eliminate five data sets as they were not fit for the purpose of our analysis because either only the source code or only the experimental data was published, but not both.
In one case, the experimental results could not be assigned to specific code snippets since they were aggregated per subject.
The data sets of two of the studies were no longer accessible and could not be retrieved by contacting the authors.
In total, out of the 22 papers with an open data set, ten could be used in our data analysis.

\subsection{Data Analysis}
\label{sec:data-analysis}

The resulting data sets provided us with a total of 427 code snippets of varying Cognitive Complexity.
In addition, the data sets contained about 24,000 individual human evaluations of the understandability of these code snippets.
To answer \textbf{RQ1} of how well the Cognitive Complexity metric is suited to measure code comprehension, we used this data and calculated correlations between Cognitive Complexity and measures of understandability.

However, the data included a variety of variables to measure understandability, namely time, correctness, subjective ratings, physiological measures, and combinations thereof.
Specifically, out of the ten studies, nine reported time, six reported correctness, four reported ratings and one reported physiological variables.
This last study provided the physiological measures in the form of the brain deactivation as a proxy of concentration levels during program comprehension gathered with an fMRI scanner~\cite{Peitek.2018}.

We do not know if these different constructs used in the studies correlate with each other in the context of code understanding, let alone if all of them are a good proxy for code comprehension.
With this discovery in mind, once we had gathered all our data, we made the decision to calculate and report the correlation values grouped by the type of understandability measure.
In this way, we could interpret the individual correlation values in their entirety in a meaningful way and were able to reach an overall conclusion.
Moreover, we introduce the following sub-questions to better represent this separation:

\begin{itemize}
    \item \textbf{RQ1.1} How do Cognitive Complexity and the time taken to understand a code snippet correlate?
    \item \textbf{RQ1.2} How do Cognitive Complexity and the percentage of correctly answered comprehension questions on a code snippet correlate?
    \item \textbf{RQ1.3} How do Cognitive Complexity and a participant's subjective ratings of a comprehension task correlate?
    \item \textbf{RQ1.4} How do Cognitive Complexity and physiological measures on a participant correlate?
    \item \textbf{RQ1.5} How do Cognitive Complexity and composite variables for code understandability correlate?
\end{itemize}

\begin{table*}[t]
	\centering
	\caption{Data sets and their properties extracted from relevant studies}
	\begin{tabular}{r r l r r r r r r l l l}
		\toprule[1pt]
		                &                           &               &               & \multicolumn{4}{|c|}{Cognitive Complexity}                     &               &                   &                           & \\
		\textbf{DID}    & \textbf{Ref}              & \textbf{Language}   & \textbf{SNo}  & \multicolumn{1}{|r}{\textbf{Med}}  & \textbf{Min}  & \textbf{Max}  & \multicolumn{1}{r|}{\textbf{SD}}   & \textbf{PNo} & \textbf{Demographic} & \textbf{Task}     & \textbf{Measures}   \\ 
		\midrule[1pt]
		 1              & \cite{Siegmund.2012}		&    	   Java &  23           & 3.0           & 0             & 9             & 2.43          & 41   & Students        & Calculate output  & Time, correctness, rating \\
		 2              & \cite{Peitek.2018} 		&        Java   &  12           & 2.0           & 0             & 6             & 1.88          & 16   & Students         & Calculate output  & Time, physiological       \\
		 3              & \cite{Buse.2010} 			&    	   Java & 100           & 1.0           & 0             & 10            & 1.74          & 121  & Students         & Rate snippet      & Rating                   \\
		 4              & \cite{Dolado.2003} 		& C\slash C++   &  20           & 2.0           & 0             & 8             & 2.35          & 51   & Prof. \& Stud.         & Answer questions  & Time, correctness         \\
		 5              & \cite{Salvaneschi.2014}   &   	  Scala &  20           & 2.0           & 0             & 7             & 1.82          & 38   & Students         & Answer questions  & Time, correctness         \\
		 6              & \cite{Scalabrino.2019} 	& 	   Java     &  50           & 7.5           & 0             & 46            & 8.53          & 63   & Professionals         & Rate and answer   & Time, correctness, rating \\
		 7              & \cite{Hofmeister.2017}	&  	    C\#     &   6           & 2.0           & 1             & 4             & 1.33          & 72   & Professionals         & Find bug          & Time                      \\
		 8              & \cite{Ajami.2017} 		&  JavaScript   &  40           & 4.0           & 1             & 14            & 3.60          & 222  & Professionals         & Calculate output  & Time                      \\
		 9              & \cite{Borstler.2016}		&  	   Java     &  30           & 4.0           & 0             & 16            & 6.81          & 259  & Students         & Rate and cloze test & Time, correctness, rating \\
		10              & \cite{Gopstein.2017}      & C\slash C++   & 126           & 1.0           & 0             & 8             & 1.40          & 48   & Students         & Calculate output  & Time, correctness         \\
		\bottomrule[1pt]
	\end{tabular}
	\label{table:data-sets-overview}
\end{table*}

\subsubsection{Data Preparation}
\label{data-preparation}

There was a significant amount of heterogeneity among the way the code snippets were provided in each study.
Not all studies provided the code snippets in source code files.
For data set 6, only the URLs to the corresponding projects that included the code snippets were provided.
Since the compilation of all of these projects proved to be difficult, just the methods themselves were extracted.
Additionally, the code snippets from data sets 1, 2, 4, 5, and 9 were provided in PDF files, so they first had to be copied into the appropriate source code files.
For data set 10, the source code was provided on the project website in text form.
These snippets were also copied into source code files.

We used SonarQube\footnote{https://www.sonarqube.org/} to analyze the source code snippets of each study as it allows us to automatically measure Cognitive Complexity.
To make this analysis work, all source code snippets had to be compilable, meaning free of syntax errors and dependency issues.
Unfortunately most of the code provided alongside the studies did not fulfill these criteria.
Additional efforts had to be made to change the snippets in ways that allowed automatic analysis without altering the Cognitive Complexity values.
These changes included, for example, adding additional import statements or libraries (data sets 1, 2, 6, 9), creating dummy classes and methods (data sets 3, 6, 9) and fixing syntax errors (data sets 3, 4).

\subsubsection{Statistical Methods and Interpretation}

\paragraph{Correlation Analysis}
Each data set contained several code snippets and usually also several measurements of code understandability.
For example, data set 5 originates from a study in which the time and correctness were measured for 20 different code snippets using 38 participants (see Table~\ref{table:data-sets-overview}).
For each code snippet, we first calculated the mean value of the measurements grouped by proxy variable.
We are aware of an open debate~\cite{Murray.2013} on whether Likert scales represent ordinal or continuous intervals.
While the debate still does not have clear indications, we opted to consider the subjective ratings using a Likert scale to be continuous in nature when calculating their mean.
Moreover, we consider Likert items as discrete values on a continuous scale.

To stay with the example of data set 5, this meant that we obtained 20 aggregated data points each for time and correctness, which serve as a proxy for the understandability of 20 code snippets.
Together with the respective values for the Cognitive Complexity of each code snippet we performed a correlation calculation.
We used Pearson's correlation coefficient where its assumptions are met, and Kendall's rank correlation coefficient (\textit{tau}) otherwise, for example, if the test for normality of the underlying data failed.
Only two of the variables from two of the data sets ended up fulfilling all of the criteria for using Pearson's correlation coefficient.
The most common reason for most of the data sets was that the cognitive complexity of the code snippets was far from a normal distribution.

On top of the data provided by the studies, additional composite variables were calculated for studies where both time and correctness of comprehension tasks were measured for the same code snippets.
We followed a similar approach to Scalabrino et al.~\cite{Scalabrino.2019} who also calculated a composite variable from time and correctness as a proxy of understandability.
In our study, the formula shown in~\cref{eq:composite} was used to calculate the composite variable \textit{timed correctness} for each snippet.
\begin{equation}
\label{eq:composite}
  (\frac{Time}{Time_{max}}) (1 - \frac{Correctness}{Correctness_{max}})
\end{equation}
Here, $Time$ and $Correctness$ refer to the median time and average correctness for this snippet.
$Time_{max}$ and $Correctness_{max}$ refer to the highest median time and average correctness calculated for all code snippets in the corresponding study.

\paragraph{Data Synthesis}
At this point, we had a set of correlation values, each of which provides information on the strength of the relationship of a proxy variable to Cognitive Complexity within a particular study.
Kendall's correlation coefficients were transformed to Pearson's r~\cite{Walker.2003} and then all correlation coefficients were converted to the Fisher's z scale for a meta-analysis using the random-effects model~\cite{Borenstein.2009.Correlations, Borenstein.2009.RandomEffects}.
This model takes into account that the true effect size of the studies varies, which can be reasonably assumed in our case.
None of the studies in their original form referred to Cognitive Complexity and even when the same proxy variable was measured, the resulting data were based on different study designs.
For that reason, we first made sure that every variable we included in the correlation analysis measures the same construct, given the definition of understandability in \cref{sec:background} of this paper.
Then we grouped the different ways of measuring code understandability into the research questions defined above.
With this approach, we believe that a combined analysis allows a meaningful interpretation and therefore contributes to answering the research questions~\cite{Borenstein.2009.Sense}.
For the calculations, we used R and the `meta' package\footnote{https://CRAN.R-project.org/package=meta} in version 4.11-0.

All correlation coefficients and their respective combined values are presented as forest plots in the following chapter.
First, such a visualization indicates a tendency, for example, whether for \textit{correctness} the majority of the data sets show a positive, negative, or no correlation with Cognitive Complexity.
Second, the summary effect represents a weighted mean and is often in line with this visually observable tendency, but can also deviate from it.
Under the random-effects model it can be seen as an estimate of the mean of the distribution of effect sizes~\cite{Borenstein.2009.RandomEffects}.
In addition to the correlation coefficients, we provide p-values but we would like to stress that they should be taken with the usual caution.
For example, some of the data sets contained only a small number of code snippets and therefore only as many data points for the correlation analysis.

\subsubsection{Data Set and Supplemental Material}
According to the principles of Open Science, we decided to publish the data~\cite{marvin_munoz_baron_2020_3949828} used in and generated by our data analysis and literature search to ensure reproducibility, repeatability, and transparency.
This includes the data for our correlation analysis and the R scripts used to calculate the correlation coefficients.
Additionally, we provide descriptions of each of the understandability studies.
Finally, we describe all the technical details of our literature search process step-by-step and list which literature was excluded in each step.
In the future, this data could be further enriched with more understandability studies and used to evaluate other source code understandability metrics.

\section{Results}
\label{sec:results}
In this chapter, we will present and describe the results of our random-effects meta-analysis of Cognitive Complexity and its correlation with measures of the understandability of source code.

Overall, we used the data of 10 studies spanning over 427 code snippets which were evaluated in terms of understandability.
The final list of data sets used to calculate correlations and their properties can be seen in \cref{table:data-sets-overview}.
The studies differed in their methodology, demographic, and materials.
The DID represents a unique identifier for each of the data sets and will be used continuing from this point on.
The table also shows some information on the code snippets provided by the study such as the programming language, the number of snippets (SNo), and the distribution of Cognitive Complexity for these snippets.
Finally, it shows some methodological information on the studies with the number of participants (PNo), the performed comprehension tasks, and which measures were taken as proxy variables for understandability.

As not all studies measured each type of proxy variable for understandability, the number of studies and snippets that could be used to answer the individual research questions \textbf{RQ1.1} to \textbf{RQ1.5} varied.
In the forest plots, the first column (\textit{Study}) describes which measurement of a study was used for the synthesis.
Variables marked as \textit{pooled} represent combined effect sizes of multiple outcomes or subgroups within a study.
The studies were either conducted with university students, professional developers, or both.
To interpret the effect sizes we used the guidelines by Cohen~\cite{Cohen.1988} and consider an effect size >0.1 to be small, >0.3 to be medium and >0.5 to be large.

\subsection*{Time variables}
\textit{\textbf{RQ1.1} How do Cognitive Complexity and the time taken to understand a code snippet correlate?}
In code understandability experiments, understandability is sometimes measured by the time taken to understand a code snippet or the time taken to complete a comprehension task on the code such as answering a question or finding a bug.
In expectation, we would assume that more difficult code snippets would take longer to understand.
For Cognitive Complexity, higher values should indicate that code is less understandable and result in longer comprehension times.
If Cognitive Complexity can accurately measure the understandability of source code, we would expect a positive correlation between Cognitive Complexity and time taken to understand code.

Nine studies were included in the analysis totaling 327 code snippets.
Overall, we observed positive results with regards to \textbf{RQ1.1} for the correlation of Cognitive Complexity and time variables.
The effect sizes ranged from a small negative correlation of $-0.03$ to a large positive correlation of $0.94$.
The weighted mean of all studies showed a large positive correlation of $0.54$.

\begin{figure}[h]
	\centering
    \includegraphics[clip, trim=0cm 4.9cm 0cm 5.4cm, width=1\linewidth]{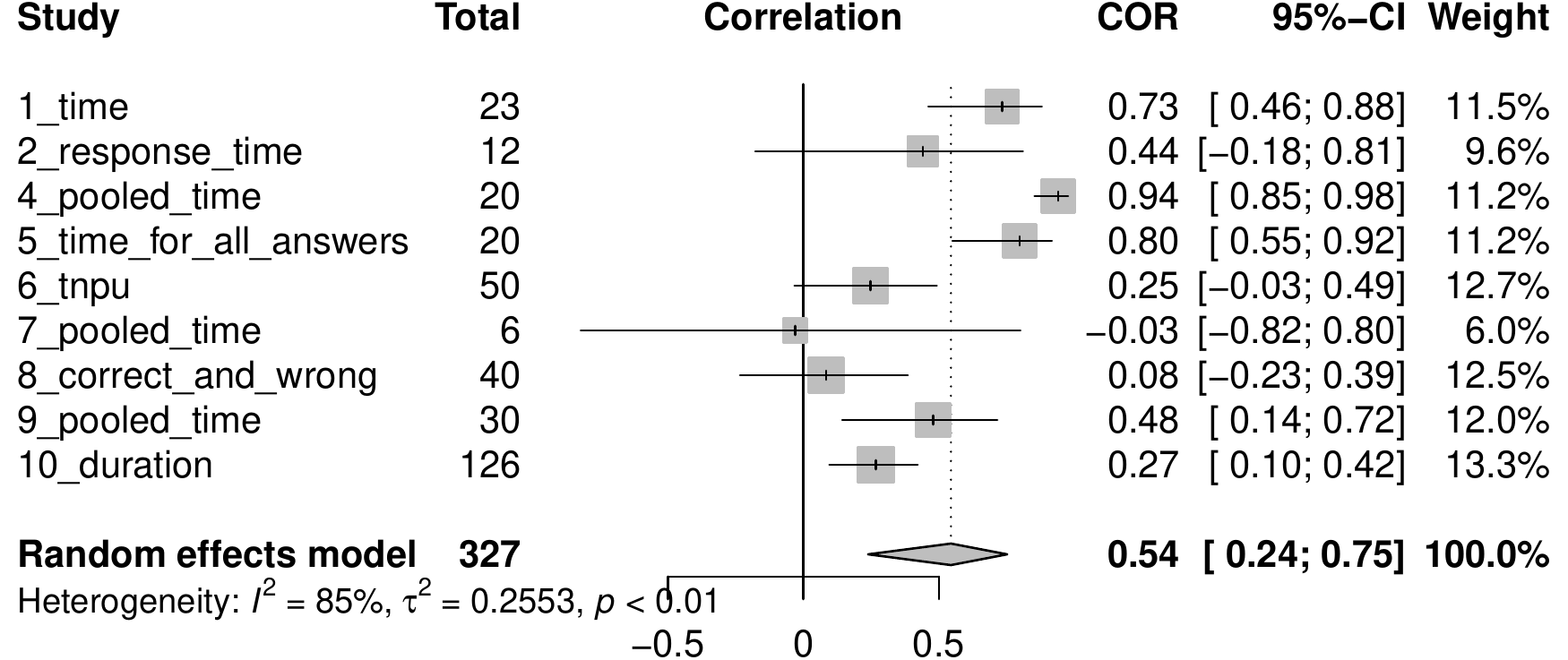}
	\caption{Forest plot for time variables}
	\label{fig:time-forestplot}
	\Description{}
\end{figure}

\subsection*{Correctness variables}
\textit{\textbf{RQ1.2} How do Cognitive Complexity and the percentage of correctly answered comprehension questions on a code snippet correlate?}
Correctness describes the degree to which the comprehension task in an experiment on source code was completed correctly.
This could, for example, be the percentage of correct answers to questions regarding code behavior or the percentage of correctly identified errors in the code.
In general, we would expect that the easier to understand a piece of code is, the higher the correctness.
For Cognitive Complexity, lower values should indicate better understandability and should result in higher correctness.
In other words, if Cognitive Complexity can accurately measure the understandability of source code, we would expect a negative correlation between Cognitive Complexity and correctness of comprehension tasks.

Six studies were included in the analysis totaling 269 code snippets.
Overall, we observed mixed results with regards to \textbf{RQ1.2} for the correlation of Cognitive Complexity and correctness variables.
The effect sizes ranged from a large negative correlation of $-0.52$ to a large positive correlation of $0.57$.
The weighted mean of all studies showed a small negative correlation of $-0.13$.

\begin{figure}[h]
	\centering
    \includegraphics[clip, trim=0cm 5.6cm 0cm 6.1cm, width=1\linewidth]{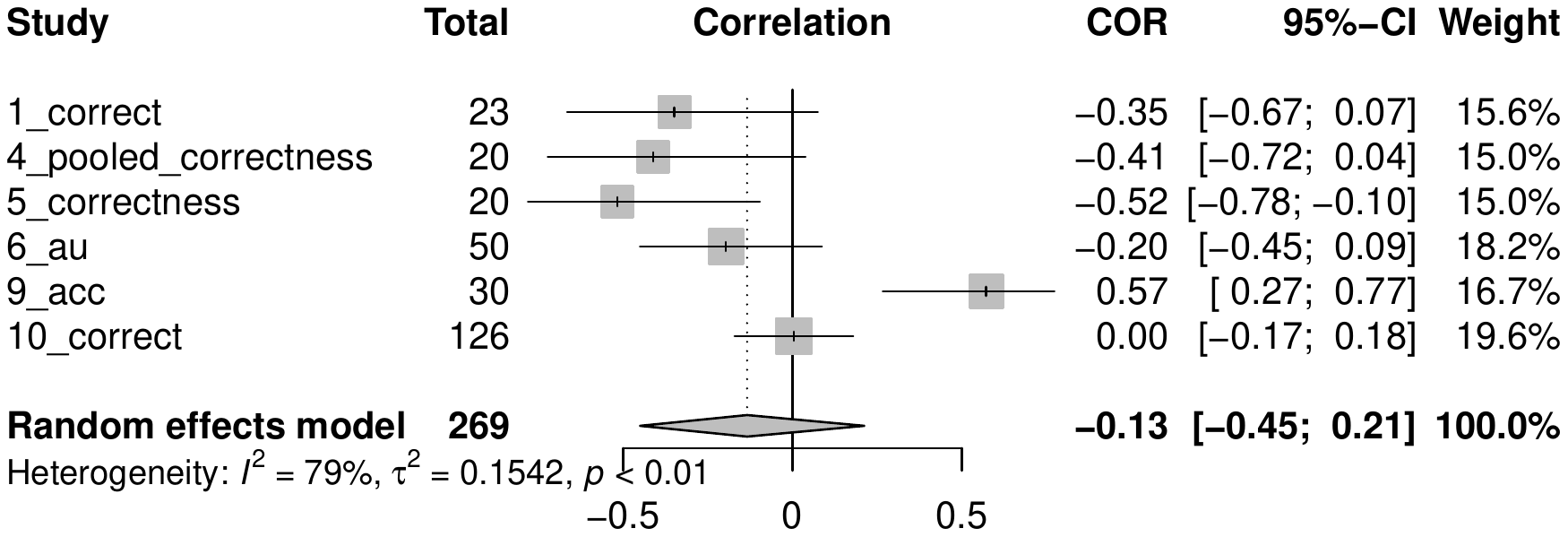}
	\caption{Forest plot for correctness variables}
	\label{fig:correctness-forestplot}
	\Description{}
\end{figure}

\subsection*{Rating variables}
\textit{\textbf{RQ1.3} How do Cognitive Complexity and a participant's subjective ratings of a comprehension task correlate?}
A rating refers to how difficult a subject believes a piece of code is to understand, sometimes called perceived understandability~\cite{Scalabrino.2019}.
For the studies included in this meta-analysis higher ratings meant that the code snippet was easier to understand.
In this case, higher values of Cognitive Complexity, as a measure of understandability, should correspond with lower ratings for code snippets.
In other words, if Cognitive Complexity can accurately measure the understandability of source code, we would expect a negative correlation between Cognitive Complexity and subjective rating of the understandability of code snippets.

Four studies were included in the analysis totaling 203 code snippets.
Overall, we observed positive results with regards to \textbf{RQ1.3} for the correlation of Cognitive Complexity and rating variables.
The effect sizes ranged from a large negative correlation of $-0.57$ to a small negative correlation of $-0.04$.
The weighted mean of all studies showed a medium negative correlation of $-0.29$.

\begin{figure}[h]
	\centering
    \includegraphics[clip, trim=0cm 6.1cm 0cm 6.6cm, width=1\linewidth]{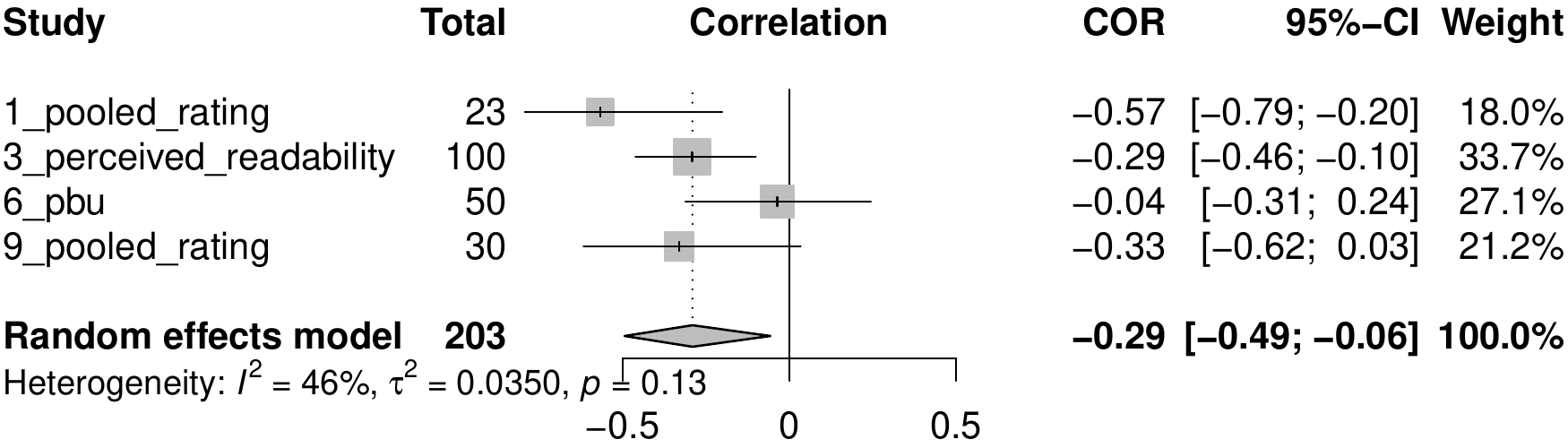}
	\caption{Forest plot for rating variables}
	\label{fig:rating-forestplot}
	\Description{}
\end{figure}

\subsection*{Physiological variables}
\begin{figure}[h]
	\centering
    \includegraphics[clip, trim=0cm 6.4cm 0cm 6.9cm, width=1\linewidth]{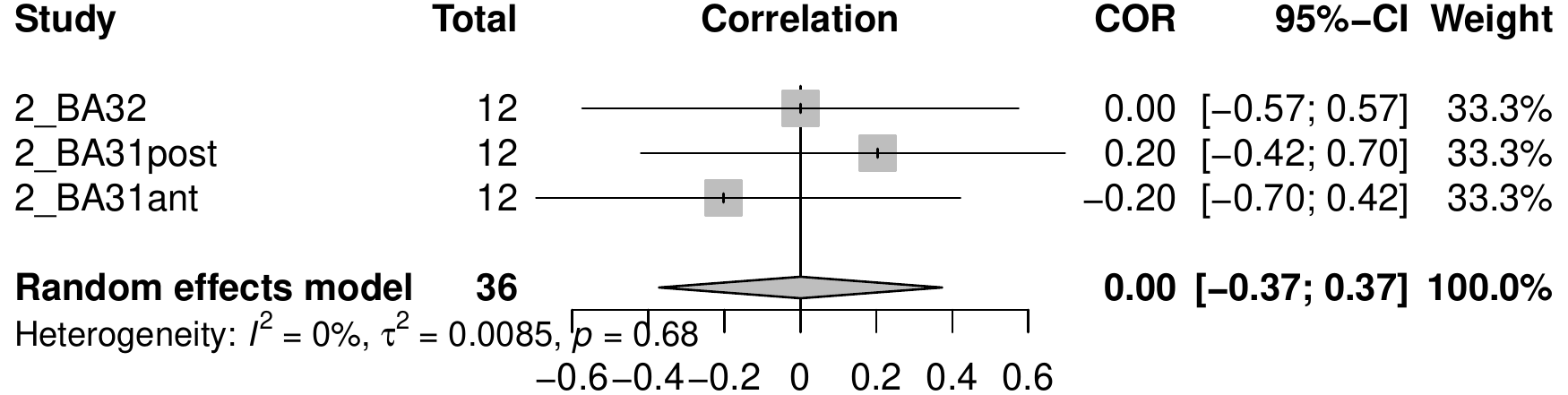}
	\caption{Forest plot for physiological variables}
	\label{fig:physiological-forestplot}
	\Description{}
\end{figure}

\textit{\textbf{RQ1.4} How do Cognitive Complexity and physiological measures on a participant correlate?}
Physiological variables are measures taken on the human body during code comprehension such as bio-metric data from EKGs or fMRIs or eye-tracking information.
In this case, the only physiological variable was the strength of brain deactivation measured as a proxy for concentration through the usage of an fMRI scanner~\cite{Peitek.2018}.
In general, we would expect less understandable snippets to require a higher level of concentration, resulting in a stronger deactivation and lower values of the physiological variable.
If Cognitive Complexity can accurately measure the understandability of source code, we would expect a negative correlation between Cognitive Complexity and the brain deactivation during comprehension of source code.
As there was only one study measuring physiological variables and providing their data, \cref{fig:physiological-forestplot} shows the results for all three physiological variables from said study.
The variables refer to different Brodmann areas (BA) of the brain.

One study was included in the analysis totaling 12 code snippets.
Overall, we observed negative results with regards to \textbf{RQ1.4} for the correlation of Cognitive Complexity and physiological variables.
For physiological variables the effect sizes ranged from a small negative correlation $-0.20$ to a small positive correlation $0.20$.
The weighted mean of all studies showed no correlation with an effect size of $0.00$.

\subsection*{Composite variables}
\textit{\textbf{RQ1.5} How do Cognitive Complexity and composite variables for code understandability correlate?}
The composite variables in this study are composed of measures of the time and correctness of the comprehension of source code snippets.
None of the composite variables were provided by the studies themselves.
Instead, they were calculated for the studies that reported both time and correctness of comprehension for the same snippets.
The formula used to calculate the composite variables can be found in~\cref{eq:composite}.
Higher values for composite variables indicate that a code snippet is harder to understand.
Therefore, for Cognitive Complexity, higher values should correspond to higher values for composite variables.
In other words, if Cognitive Complexity can accurately measure the understandability of source code, we would expect a positive correlation between Cognitive Complexity and composite variables for the understandability of code snippets.

Six studies were included in the analysis totaling 269 code snippets.
Overall, we observed positive results with regards to \textbf{RQ1.5} for the correlation of Cognitive Complexity and composite variables.
The effect sizes ranged from a small negative correlation of $-0.10$ to a large positive correlation of $0.68$.
The weighted mean of all studies showed a medium positive correlation of $0.40$.

\begin{figure}[h]
	\centering
    \includegraphics[clip, trim=0cm 5.6cm 0cm 6.1cm, width=1\linewidth]{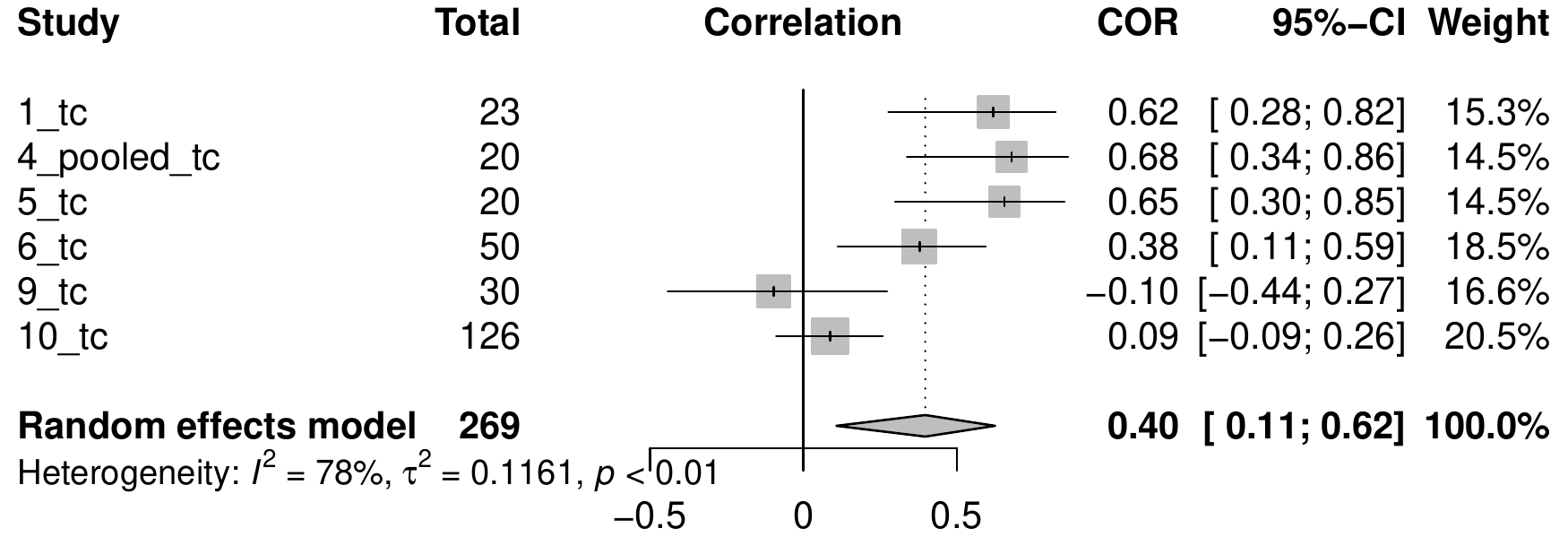}
	\caption{Forest plot for composite variables}
	\label{fig:composite-forestplot}
	\Description{}
\end{figure}

\section{Discussion}
\label{sec:discussion}

We discuss the results, limitations and implications of the study.

\subsection{Findings}

With respect to the results of our meta-analysis, we found that Cognitive Complexity correlates with the time taken to comprehend code snippets ($0.54$) and with the perceived understandability ratings from developers ($-0.29$).
When comparing these results with previous studies, Scalabrino et al.~\cite{Scalabrino.2019} found in their study of code metrics that for time variables, only 1 out of 121 metrics had a weak correlation of $0.11$ and only 8 out of 121 had a weak correlation with ratings, with the highest being $-0.13$.
Other studies have also reported that in some cases there is no correlation between different software metrics and time~\cite{Feigenspan.2011, Jbara.2014}.
When comparing the results it has to be kept in mind that these studies report the results of a single experiment, while our results pertain to combined effect sizes of multiple studies.

Our results showed little to no support for the assumption that Cognitive Complexity correlates with the correctness of code comprehension tasks.
Other studies have shown similar results for other source code metrics~\cite{Feigenspan.2011, Jbara.2014, Scalabrino.2019}.
This does not, however, imply that each code snippet was equally easy to understand.
It is likely that while participants experienced more difficulties with some snippets, this was not reflected in the values for correctness, as they may have compensated for this by spending more time with the snippet to ensure their answers were correct.
This concept is also mentioned by Borstler et al.~\cite{Borstler.2016} who noticed that negative impacts of snippets sizes could be compensated by subjects by spending more time on larger snippets.
Correlations with time and subjective ratings but not correctness suggest that there might be a correlation with the cognitive load required for understanding a piece of code.
Lastly, when combining the values for time and correctness of the same code snippet, we again observed similar results to time variables, with a combined effect size of $0.40$, which is slightly lower than the $0.54$ for time variables.
Physiological measures could play an important role in finding an answer to these questions as they might be good indicators of the cognitive load of code comprehension tasks.

With the given data in our study, we could not identify a correlation between Cognitive Complexity and the brain deactivation during code comprehension.
When comparing the results of our study to Peitek et al.~\cite{Peitek.2018}, the source of data for our study, we find similar results.
While Cognitive Complexity performs slightly better than Cyclomatic Complexity in BA31post region, the correlations are overall small and insignificant.
Peitek et al. themselves mention that these results should be taken with a grain of salt as the sample size and number of code snippets was quite small.
Regardless, both analyses show possibilities as to how physiological measures can be used in understandability studies to investigate correlations with software metrics.
Larger data sets might reveal more conclusive evidence as to how Cognitive Complexity or other metrics correlate with physiological measures such as the strength of brain deactivation during code comprehension.

To answer \textbf{RQ1}, we conclude that with respect to the results of this analysis, Cognitive Complexity is the first validated and solely code-based metric that is able to reflect at least some aspects of code understandability.
While there are some mixed results for the correctness and physiological variables, the results for time, composite and rating variables show empirical support for Cognitive Complexity as a measure of source code understandability.
Both in terms of industry-appliance and research contexts, being able to estimate the time it takes a developer to understand a source code snippet as well as how they might perceive it in terms of understandability can be of great value.
Cognitive Complexity appears to be a metric that can achieve these goals based on the empirical evidence provided by this study.

Our systematic literature search revealed a multitude of insights into the way understandability studies are conducted.
We found that there a many different ways researchers define, measure and correlate understandability.
When attempting to summarize and combine the results from different studies and different measures, our efforts ultimately revealed that studies on source code comprehension require a uniform understanding of the studied construct as well as comparable and validated measures.
While it is a tempting thought to combine the types of variables rather than dividing each of them into separate research questions, in our case the lack of information on the relationships and validity of these measures prevented us from doing so.
Without sound reasoning and thorough scientific analysis, there is no conclusive evidence as to how these measures correlate with each other and furthermore which of them are a valid way to express code understanding.
We have to address this issue in the future to make the results comparable and to find a validated way of measuring code comprehension.

With the results of our literature search we created an aggregated data set of measurements of the Cognitive Complexity and understandability of code snippets.
In line with the studies we extracted data from, we openly publish our data and methods to ensure reproducibility and transparency.
While we used this aggregated data set to evaluate the Cognitive Complexity metric, future works could expand on our work and use the data set in order to measure the correlations of other metrics with proxy variables of understandability.
Unfortunately we were not able to republish the code snippets alongside the data we used to calculate the correlations, as we were not able to obtain licenses of the published code.
We did, however, provide a link to the code snippets and their unique identification for each of the data points.
While the data provided by the studies was in good shape, publishing code snippets under a specific license is still something that is not common.

Around 40\% of relevant studies openly publish their data sets.
While there is still room for improvement, we can observe a positive trend with regards to open accessibility of scientific data, especially in recent years.
Overall, collecting data from existing studies in an effort to validate a metric allowed us to have a much larger sampling size in comparison to similar studies.
Using preexisting data sets also comes with some difficulties, as searching and aggregating experimental data and preparing code snippets to gather metrics can be quite complex and time consuming.
Other limitations with the approach used in this study are discussed in more detail in the following subsection~\ref{sec:limitations}.

\subsection{Limitations}
\label{sec:limitations}

Our work consists of two main parts: the search for relevant existing data sets in which code snippets have been evaluated in their understandability and the synthesis of these data for the purpose of validating a metric.
The results of both parts should be seen in the light of some limitations.

We started the systematic literature search with database searches.
The search term had to be broad and therefore produced a very large number of irrelevant search results that had to be filtered manually.
This process can be error-prone.
We have countered this by having one author perform the filtering and a second one validating the results.
In the future a comparable systematic literature search could benefit from a pure snowballing approach, where less irrelevant results are suspected~\cite{Wohlin.2014}.

We decided to limit the initial results of the literature search to publications published after the year 2010 as we suspected those to most likely have an open data set provided alongside their study or their authors be most likely to share their data with us.
This possibly introduced a sampling bias as more recent studies had a higher probability of being included in the search results.
We attempted to alleviate this limitation by additionally including a snowballing step without the publishing period restriction.

Then we had to search the texts of all studies in question for a reference to their data set, whereby we might have missed references.
Unfortunately, there is no publisher-independent uniform system or paper meta-data for identifying supplemental material.
This would have allowed us to efficiently access the corresponding data record.
Many publishers do not reference supplemental material at all.
However, missed data set references are not an issue, as in such cases we have contacted the authors of relevant studies to request access to the code snippets and experimental data.

Additionally, some code snippets had to be altered in order to free them of syntax errors and dependency issues so that their Cognitive Complexity could be calculated automatically by SonarQube.
This could potentially represent a threat to the validity of the study, as altering the code snippets manually might change their value for Cognitive Complexity.
In order to mitigate this, we made sure to only make changes that would not influence the Cognitive Complexity of a snippet, such as adding closing brackets or fixing dependency issues.
None of the 427 snippets had to be changed in a way that would have introduced new control-flow structures.
The alternative to this solution would have been to manually calculate the Cognitive Complexity for each of the snippets.
While this would have eliminated the need for altering the code, it would also have introduced a large amount of additional manual labor, increasing the risk for human error.
Due to the large number of snippets and the aforementioned conditions, we chose to go with the first solution.

Then we came to the actual analysis part, which can be described in summary as meta-analysis using the random-effects model and effect sizes based on correlations.
First, it is not a meta-analysis in the classical sense, which is due to the fact that most underlying studies did not answer the same research questions as we did.
In other words, most of the synthesized studies did not focus on the validation of metrics, and none focused on the metric that our paper is about.
This is not a limitation in itself, one only has to be aware that the original studies had different objectives and therefore varied in their designs.
We made sure that the measured data aimed at the same construct of code understandability and we used the random-effects model, which takes into account that the true effect size of the studies varies due to the different study designs and sample characteristics.
How well the respective study designs and collected data reflect code understanding in practice, however, is a threat to external validity and a question that we cannot answer in the scope of this work.
At least one positive aspect is that the data comes from many different contexts, which applies to the reality in which code must be understood in different scenarios.

Finally, few of the original studies focused on the evaluation of source code metrics, which meant that most code snippets were not necessarily intended for a comprehensive metric evaluation.
Although our analysis is based on a large number of 427 different code snippets from different languages and projects, almost all of them were of relatively low Cognitive Complexity.
Only two of the studies included code snippets with a value greater than 15, which is the default threshold in SonarQube for reporting on \textit{too complex} functions.
The correlations that were found are still meaningful, but we have little information about how well the metric works for higher values of Cognitive Complexity and accordingly no recommendation for a meaningful threshold.

\subsection{Implications}

We found Cognitive Complexity to be a good indicator of how long it takes developers to understand a piece of code and we found the composition of time and correctness to correlate positively with understandability. 
In addition, code snippets with higher Cognitive Complexity are also rated as more difficult to understand.
Since developers already spend a large part of their time trying to understand code, it is worthwhile to reduce this effort by simplifying code sections that are difficult to understand.
The metric helps to point out such sections.
Unfortunately, the question remains open as to when a section of code can actually be described as \textit{too complex}.
Future work will have to investigate what an appropriate threshold value could be. Until then the recommendation is to keep Cognitive Complexity of a code snippet as low as possible.

By validating the metric, we realized that the measurement of code understandability is still in its very beginnings.
The comparability of results, let alone their synthesis, requires a better understanding of the different ways to measure code understanding.
When understanding is measured in different ways, such as processing time and correctness of answers to comprehension questions, we need to understand how they are related and what measures are actually appropriate in which context.
Otherwise, researchers will continue to have difficulties in developing a suitable design for their code understanding experiment~\cite{Siegmund.2016}, and we will not know afterwards which findings are actually comparable.
Then we have no choice but to do what we did and, for example, break down the validation of a metric into five different research questions.

\section{Conclusion}
\label{sec:conclusion}

Understanding code is an essential part of the developer's day-to-day work.
It is therefore important to write source code in such a way that the time and mental effort required for understanding it is as low as possible.
We showed that Cognitive Complexity is a promising metric for automatically measuring different facets of code comprehension, making it easier to identify sections of code that are hard to understand.
The metric correlates with the time it takes a developer to understand source code, with a combination of time and correctness, and with subjective ratings of understandability.
This means that Cognitive Complexity is the first validated and purely code-based metric that can measure code understandability.
Although we do not know at what metric value a section of code can be considered \textit{too} complex, our recommendation is to keep the metric value as low as possible to reduce the effort of understanding source code.

To validate the metric, we first conducted a systematic literature search to find data sets that evaluate source code by human participants for its understandability.
We then synthesized the data sets in a meta-study to investigate whether Cognitive Complexity correlates with the evaluations provided by the data sets.
Since different studies measure code understandability in different ways, we divided the synthesis into different research questions to summarize the data in a meaningful way.
The steps and intermediate results of the literature search as well as all data generated during the analysis are made available in a public data set.
This enables not only reproducibility and repeatability, but also the validation of other code understanding metrics in the future.

For the metric of Cognitive Complexity itself, a meaningful threshold value has yet to be identified. 
We also know too little about how well the metric reflects physiological measures in the context of code understanding.
The number of studies carrying out such measurements is growing, so the resulting data could be used for further validation in the near future.
Finally, this study already provided us with exciting discoveries by revealing that currently, source code understandability is measured in many different ways with no clear agreement as to how those ways relate to each other.
This motivates us to expand the systematic literature search into a more comprehensive review to gain further insights into the investigated research questions, the measures used, and the difficulties in designing understandability experiments.
In this way, our long-term goal is to make the results of code comprehension studies more comparable and allow others to conduct meta-studies like the present one.


\bibliographystyle{ACM-Reference-Format}
\bibliography{ms}

\end{document}